\PassOptionsToPackage{unicode}{hyperref}
\PassOptionsToPackage{hyphens}{url}
\PassOptionsToPackage{dvipsnames,svgnames,x11names}{xcolor}
\documentclass[
  10pt,
]{article}
\usepackage{xcolor}
\usepackage{amsmath,amssymb}
\usepackage{iftex}
\ifPDFTeX
  \usepackage[T1]{fontenc}
  \usepackage[utf8]{inputenc}
  \usepackage{textcomp} 
\else 
  \usepackage{unicode-math} 
  \defaultfontfeatures{Scale=MatchLowercase}
  \defaultfontfeatures[\rmfamily]{Ligatures=TeX,Scale=1}
\fi
\usepackage{lmodern}
\ifPDFTeX\else
\fi
\IfFileExists{upquote.sty}{\usepackage{upquote}}{}
\IfFileExists{microtype.sty}{
  \usepackage[]{microtype}
  \UseMicrotypeSet[protrusion]{basicmath} 
}{}
\makeatletter
\@ifundefined{KOMAClassName}{
  \IfFileExists{parskip.sty}{%
    \usepackage{parskip}
  }{
    \setlength{\parindent}{0pt}
    \setlength{\parskip}{6pt plus 2pt minus 1pt}}
}{
  \KOMAoptions{parskip=half}}
\makeatother
\usepackage{longtable,booktabs,array}
\usepackage{calc} 
\usepackage{etoolbox}
\makeatletter
\patchcmd\longtable{\par}{\if@noskipsec\mbox{}\fi\par}{}{}
\makeatother
\IfFileExists{footnotehyper.sty}{\usepackage{footnotehyper}}{\usepackage{footnote}}
\makesavenoteenv{longtable}
\NewDocumentCommand\citeproctext{}{}
\NewDocumentCommand\citeproc{mm}{%
  \begingroup\def\citeproctext{#2}\cite{#1}\endgroup}
\makeatletter
 \let\@cite@ofmt\@firstofone
 \def\@biblabel#1{}
 \def\@cite#1#2{{#1\if@tempswa , #2\fi}}
\makeatother
\newlength{\cslhangindent}
\newlength{\csllabelwidth}
\newenvironment{CSLReferences}[2] 
 {\begin{list}{}{%
  \setlength{\itemindent}{0pt}
  \setlength{\leftmargin}{0pt}
  \setlength{\parsep}{0pt}
  \ifodd #1
   \setlength{\leftmargin}{\cslhangindent}
   \setlength{\itemindent}{-1\cslhangindent}
  \fi
  \setlength{\itemsep}{#2\baselineskip}}}
 {\end{list}}
\usepackage{calc}

\newcommand{\CSLLeftMargin}[1]{\parbox[t]{\csllabelwidth}{\strut#1\strut}}
\newcommand{\CSLRightInline}[1]{\parbox[t]{\linewidth - \csllabelwidth}{\strut#1\strut}}

\usepackage[letterpaper,margin=0.65in]{geometry}
\usepackage{graphicx}
\usepackage{caption}
\usepackage{needspace}
\usepackage{titlesec}
\usepackage{placeins}
\titleformat{\section}{\large\bfseries}{}{0pt}{}
\titleformat{\subsection}{\normalsize\bfseries}{}{0pt}{}
\titlespacing*{\section}{0pt}{10pt plus 2pt minus 1pt}{5pt}
\titlespacing*{\subsection}{0pt}{8pt plus 1pt}{4pt}
\usepackage{bookmark}
\IfFileExists{xurl.sty}{\usepackage{xurl}}{} 
\hypersetup{
  colorlinks=true,
  linkcolor={blue},
  filecolor={Maroon},
  citecolor={Blue},
  urlcolor={blue},
  pdfcreator={LaTeX via pandoc}}

\author{}
\date{}

\begin{document}

\begin{center}
{\Large\bfseries Crystoria: A Large-Scale Crystal Dataset Linking Thermodynamic and
Dynamical Stability for Materials AI\par}
\end{center}

Ke-Xin Pang\(^{1}\), Ze-Feng Gao\(^{2,3,*}\), Yan Gao\(^{1,\dagger}\),
and Zhong-Yi Lu\(^{2,3,\ddagger}\)

\(^{1}\)State Key Laboratory of Metastable Materials Science and
Technology and Hebei Key Laboratory of Microstructural Material Physics,
School of Science, Yanshan University, Qinhuangdao 066004, China\\
\(^{2}\)School of Physics and Key Laboratory of Quantum State
Construction and Manipulation (Ministry of Education), Renmin University
of China, Beijing 100872, China\\
\(^{3}\)AI-Driven Quantum Materials Research Center, School of Physics,
Renmin University of China

\(^{*}\)Correspondence: zfgao@ruc.edu.cn\\
\(^{\dagger}\)Correspondence: yangao9419@ysu.edu.cn\\
\(^{\ddagger}\)Correspondence: zlu@ruc.edu.cn

Article type: Data Express Letter

\subsection{Abstract}\label{abstract}

Large-scale crystal databases have transformed materials discovery by
making millions of structures and thermodynamic properties accessible to
materials informatics. Yet dynamical stability remains sparsely
characterized at a comparable scale because its evaluation is
substantially more computationally demanding than conventional structure
and energy calculations. This mismatch creates a critical gap between
the rapidly expanding crystal-structure space and the physical
information available for materials research and machine learning. Here
we introduce Crystoria, a large-scale crystal dataset designed to bring
dynamical stability to the scale of modern materials data. We construct
Crystoria by systematically curating low-energy and metastable inorganic
crystals, standardizing their structural identities, and combining
machine-learning-accelerated structural relaxation with phonon-based
dynamical screening. The resulting dataset contains 2,649,640 crystal
structures, spanning 89 elements, 219 of the 230 space groups, and
primitive cells containing 1--48 atoms. Among them, 1,793,353 structures
(67.68\%) are classified as dynamically stable, revealing substantial
variation in dynamical stability across chemical and crystallographic
space. By bringing dynamical stability to the scale of millions of
crystal structures, Crystoria provides a physically grounded foundation
for materials artificial intelligence (AI) and enables systematic
exploration of the dynamical crystal landscape.

Keywords: crystal structures; thermodynamic stability; dynamical
stability; phonons; machine-learning interatomic potentials; materials
informatics; scientific datasets

\subsection{Specifications Table}\label{specifications-table}

Table 1. Data specifications for Crystoria.

{\def\LTcaptype{none} 
\begin{longtable}[]{@{}
  >{\raggedright\arraybackslash}p{(\linewidth - 2\tabcolsep) * \real{0.2000}}
  >{\raggedright\arraybackslash}p{(\linewidth - 2\tabcolsep) * \real{0.8000}}@{}}
\toprule\noalign{}
\begin{minipage}[b]{\linewidth}\raggedright
Item
\end{minipage} & \begin{minipage}[b]{\linewidth}\raggedright
Specification
\end{minipage} \\
\midrule\noalign{}
\endhead
\bottomrule\noalign{}
\endlastfoot
Subject & Materials physics \\
Specific subject area & Crystal thermodynamics and lattice dynamics \\
Type of data & Atomic structures, hull energies, phonon frequencies and
stability classes \\
Data format & Processed computational data from crystal curation, model
relaxation and phonon calculations; Parquet tables with structured
atomic coordinates \\
Data source location & Alexandria, Materials Project, JARVIS, NIMS
HeuslerDB and OQMD; some Alexandria and Materials Project data obtained
through MatterGen \\
Data accessibility & Zenodo (version 1.0.0); \href{https://doi.org/10.5281/zenodo.23572554}{\textcolor{black}{10.5281/zenodo.23572554}} \\
Related research & None \\
\end{longtable}
}

\Needspace{6\baselineskip}

\subsection{1. Background \& Summary}\label{background-summary}

The rapid growth of high-throughput computation and data-driven methods
has fundamentally changed the way crystalline materials are discovered
and studied. Large computational resources, including the Materials
Project, AFLOW, OQMD, JARVIS, NOMAD, and Materials Cloud, have
established a data-centric paradigm in which crystal structures and
calculated materials properties can be systematically explored across
broad chemical
spaces\textsuperscript{\citeproc{ref-Jain2013MaterialsProject}{1}--\citeproc{ref-Talirz2020MaterialsCloud}{6}}.
The development of interoperable computational infrastructures and open
materials-analysis tools has further enabled the generation,
organization, provenance tracking, and reuse of large-scale
computational materials
data\textsuperscript{\citeproc{ref-Pizzi2016AiiDA}{7}--\citeproc{ref-Ramprasad2017MaterialsInformatics}{10}}.
Large-scale thermodynamic data have in turn enabled systematic analyses
of inorganic stability and metastability, establishing energy above the
convex hull as a central descriptor for computational materials
discovery\textsuperscript{\citeproc{ref-Sun2016Metastability}{11}--\citeproc{ref-Bartel2020StabilityML}{13}}.
More recently, machine-learning approaches have shifted materials
research from fitting properties of individual compounds toward learning
transferable representations and exploring increasingly large structural
spaces\textsuperscript{\citeproc{ref-Ward2016MLMaterials}{14}--\citeproc{ref-Schmidt2019MLSolidState}{16}}.
Large-scale data-driven discovery has now reached a qualitatively
different regime: GNoME reported more than two million structures
predicted to be stable with respect to previous computational data,
while MatterGen demonstrated generative design of diverse inorganic
crystals using a large curated corpus of stable
structures\textsuperscript{\citeproc{ref-Merchant2023GNoME}{17},\citeproc{ref-Zeni2025MatterGen}{18}}.
The resulting expansion of accessible crystal space has made the
quality, diversity, and physical content of materials data increasingly
important for both computational discovery and materials artificial
intelligence
(AI)\textsuperscript{\citeproc{ref-Curtarolo2013HTHighway}{19}--\citeproc{ref-Jain2016EnergyMaterials}{21}}.
Building large-scale materials datasets with physically meaningful
property information is therefore becoming a central requirement for the
next generation of data-driven materials research.

Against this rapid expansion of crystal-structure data, efforts to
characterize dynamical stability have also progressed, but on a
substantially smaller scale. First-principles lattice-dynamical
calculations provide a direct computational route to determining whether
a crystal is locally stable against collective atomic
displacements\textsuperscript{\citeproc{ref-Togo2015Phonons}{22},\citeproc{ref-Gonze1997DFPT}{23}}.
Early high-throughput work demonstrated the feasibility of
systematically evaluating phonon properties for inorganic materials,
including complete density-functional perturbation-theory phonon
calculations for 1,521
compounds\textsuperscript{\citeproc{ref-Petretto2018Phonons}{24}}.
Dedicated phonon repositories subsequently expanded the accessible
collection to approximately \(10^4\) harmonic phonon calculations,
providing a standardized source of dynamical information for inorganic
crystals\textsuperscript{\citeproc{ref-TogoMDRPhonon}{25}}. In parallel,
high-throughput studies have used phonon calculations to screen
dynamically stable materials within targeted chemical families,
demonstrating the value of incorporating lattice dynamics into
large-scale materials
searches\textsuperscript{\citeproc{ref-Lee2025HighThroughputPhonons}{26},\citeproc{ref-Xiao2025HeuslerPhonons}{27}}.
The emergence of machine-learning interatomic potentials (MLIPs) has
opened a further route toward scaling dynamical calculations beyond what
is practical with direct first-principles
evaluation\textsuperscript{\citeproc{ref-Behler2007NNPotential}{28}--\citeproc{ref-Yang2024MatterSim}{34}}.
Recent work has begun to systematically benchmark universal MLIPs on
phonon properties using approximately 10,000 first-principles phonon
calculations, showing that several modern models can reproduce harmonic
phonon properties with high
accuracy\textsuperscript{\citeproc{ref-Loew2025Phonons}{35}}. This
progression from direct first-principles calculations to
machine-learning-accelerated phonon evaluation has recently enabled
dynamical-stability datasets and benchmarks at the \(10^5\) scale,
including PhononBench-MP40 with 46,899 completed phonon records and
PhononBench with 133,838 AI-generated crystals evaluated for dynamical
stability\textsuperscript{\citeproc{ref-Li2026PhononBenchMP40}{36},\citeproc{ref-Han2026PhononBench}{37}}.
Nevertheless, these resources remain substantially smaller than modern
general-purpose crystal databases, which already contain millions of
structures\textsuperscript{\citeproc{ref-Schmidt2024Alexandria}{38},\citeproc{ref-Cavignac2025AlexandriaExpansion}{39}}.
Thus, although the scale of dynamical-stability characterization has
increased by orders of magnitude, a pronounced gap remains between the
size of available crystal-structure spaces and the number of structures
for which systematic dynamical-stability data are available.

Recent advances in machine-learning interatomic potentials have made
this scale gap increasingly tractable. Modern universal MLIPs are
trained on large and chemically diverse first-principles datasets and
use rotationally equivariant representations to improve the description
of atomic interactions across broad chemical
spaces\textsuperscript{\citeproc{ref-Chen2022M3GNet}{30}--\citeproc{ref-Yang2024MatterSim}{34},\citeproc{ref-Wood2025UMA}{40}}.
Importantly, their suitability for dynamical calculations is no longer
purely prospective. The phonon benchmarks also reveal model-dependent
failures that are not necessarily apparent from energy and force tests
alone. These differences make direct vibrational comparisons an
essential part of assessing a model for large-scale dynamical-stability
screening\textsuperscript{\citeproc{ref-Loew2025Phonons}{35},\citeproc{ref-Deng2025Softening}{41}}.
At the level of large-scale materials discovery, the current generation
of force fields has also reached high task-level performance. In the
Matbench Discovery benchmark, which evaluates machine-learning models in
a prospective high-throughput materials-discovery setting, EquFlashV2
achieves a thermodynamic-stability-classification accuracy of
approximately \(97.8\%\) and an F1 score of \(0.929\), with a combined
performance score of
\(0.907\)\textsuperscript{\citeproc{ref-Riebesell2025Matbench}{42},\citeproc{ref-EquFlashV2}{43}}.
These results establish that contemporary MLIPs can provide both the
atomistic fidelity and computational efficiency needed for large-scale
structural relaxation and stability screening. The remaining challenge
is therefore not simply the availability of accurate atomistic models,
but how their capabilities can be systematically deployed to generate
consistent and trustworthy dynamical data across a rapidly expanding
crystal-structure space. This requires coordinated treatment of
structural relaxation, consistent recognition of equivalent crystal
structures, the source and calculation conditions of thermodynamic data,
phonon-based classification, near-zero imaginary modes, unsuccessful
calculations, and independent higher-fidelity validation. The central
question is consequently: how can reliable dynamical-stability data be
generated at the scale of modern crystal databases while maintaining
physical consistency, reproducibility, and traceability?

Here we address this challenge by introducing Crystoria, a large-scale
crystal dataset designed to bring dynamical-stability characterization
to the scale of modern materials data. We systematically curate
low-energy and metastable inorganic crystal structures from multiple
thermodynamic sources, standardize their structural representations and
identities, and integrate machine-learning-accelerated structural
relaxation with a uniform phonon-based dynamical-stability screening
workflow. The phonon force model is assessed against external
density-functional-theory (DFT) phonon calculations to test the
reliability of its dynamical classifications. Crystoria contains
2,649,640 crystal structures, spanning 89 elements, 219 of the 230 space
groups, and primitive cells containing 1--48 atoms. Among them,
1,793,353 structures (67.68\%) are classified as dynamically stable,
providing dynamical-stability data across broad chemical and
crystallographic spaces. By extending systematic dynamical-stability
characterization from specialized datasets in the \(10^5\) regime to the
million-structure scale, Crystoria establishes a large-scale, physically
curated resource for incorporating dynamical stability into
computational materials discovery and materials artificial intelligence.

\subsection{2. Methods}\label{methods}

Extending dynamical-stability information to millions of crystals
requires each phonon result to refer to a well-defined atomic
arrangement. We bring together primitive cells from Alexandria,
Materials Project, JARVIS, NIMS HeuslerDB, and OQMD, retaining their
original structure and calculation sources (Fig. 1a). Some Alexandria
and Materials Project structures and reference energies are obtained
through the data collections distributed with
MatterGen\textsuperscript{\citeproc{ref-Zeni2025MatterGen}{18}}. The
same periodic crystal can appear under different lattice bases, origins,
or database entries. The three TiCoSb representations illustrate how
such descriptions are reduced to a common primitive-cell representation
before repeated structures are removed. This step makes the subsequent
energy and phonon calculations refer to distinct crystal arrangements
rather than alternative descriptions of the same lattice. Spatial
inversion is treated separately because it can reverse handedness;
explicitly generated inversion partners are retained. Complete
construction procedures are given in Supplementary Methods
S1\textsuperscript{\citeproc{ref-CrystoriaSupplementaryInformation}{44}}.

\begin{figure}[!htbp]
\centering
\includegraphics[width=\linewidth,height=0.67\textheight,keepaspectratio]{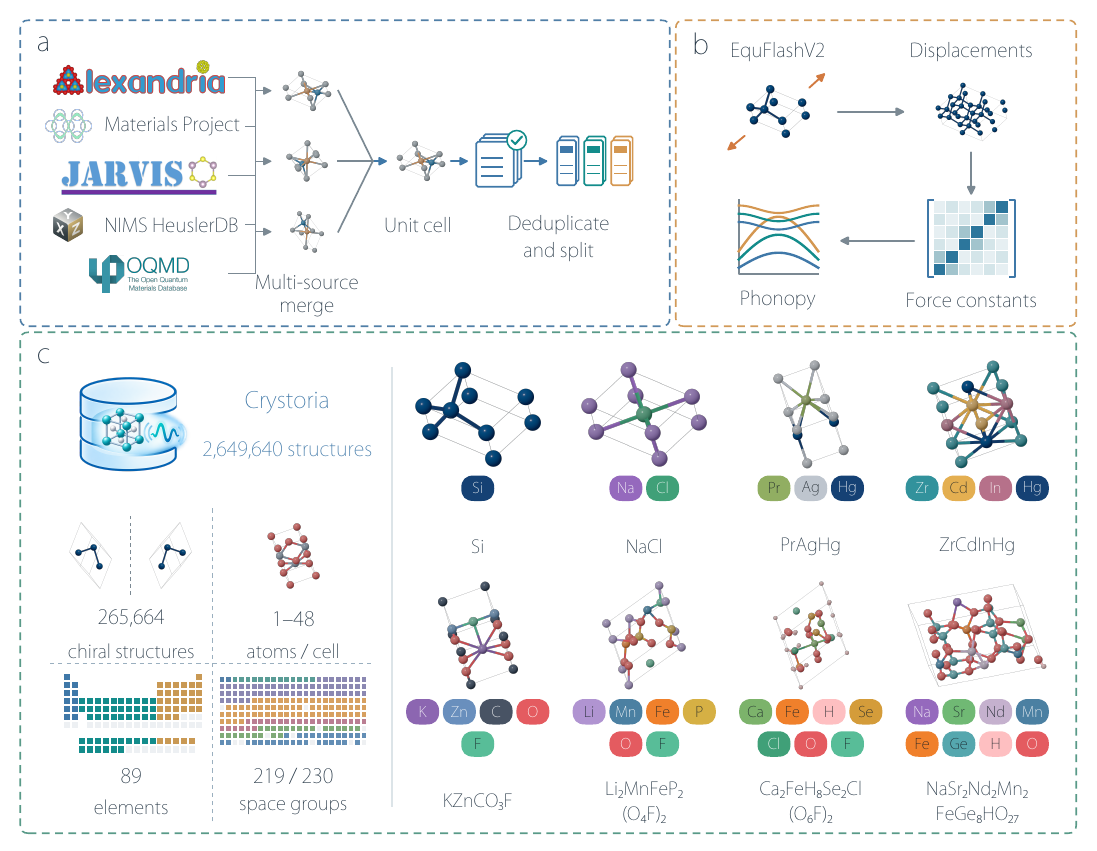}
\caption{From crystal structures to dynamical-stability data. (a) Five database
sources are brought into a common structural description. Equivalent
TiCoSb cells represent one crystal; explicit inversion partners are
retained and linked to their calculation sources. The source-energy
selection is \(E_{\mathrm{hull}}\leq100\) meV atom\(^{-1}\). (b)
EquFlashV2 lattice relaxation, with a 0.03 eV Å\(^{-1}\) force
criterion, followed by 0.01 Å displacements in a \(2\times2\times2\) Si
supercell and phonon evaluation on a \(12\times12\times12\) mesh. (c)
The 2,649,640 structures include 265,664 identified as chiral in their
source cells. Te illustrates inversion-related geometry, and
\(\mathrm{SnO_2}\) illustrates primitive-cell size. Eight relaxed
structures satisfying the stable-class margin span one to eight
elements. Filled element keys identify atomic species; space-group tiles
distinguish crystal systems. Crystal structures are rendered from their
atomic coordinates with MatterViz.}
\label{fig:1}
\end{figure}

Starting structures are ordered, three-dimensional crystals with 1--48
atoms per primitive cell and source hull energies no greater than 100
meV atom\(^{-1}\). Hull energy measures competition with other phases at
the same overall composition, restricting the collection to low-energy
and metastable candidates. Where several admissible source values exist,
the largest controls selection before relaxation. Primitive-cell
reduction and symmetry analysis use pymatgen and
spglib\textsuperscript{\citeproc{ref-Ong2013Pymatgen}{8},\citeproc{ref-togo2024spglib}{45}},
with species-preserving structural comparisons.

The selected structures then undergo a common sequence of relaxation and
lattice-dynamical evaluation, illustrated by the Si cell and displaced
supercell in Fig. 1b. EquFlashV2 relaxes both atomic positions and the
cell until the maximum residual force reaches the 0.03 eV Å\(^{-1}\)
criterion, within a limit of 200
steps\textsuperscript{\citeproc{ref-EquFlashV2}{43}}. Atomic
displacements of 0.01 Å are applied in \(2\times2\times2\) supercells,
and the induced forces determine how a displacement of one atom changes
the forces on its neighbours. These couplings form the force-constant
matrix. Phonopy uses the symmetrized force constants to obtain
collective vibrational modes on a non-\(\Gamma\)-centred
\(12\times12\times12\)
mesh\textsuperscript{\citeproc{ref-Togo2015Phonons}{22}}. The
calculation therefore proceeds from a relaxed atomic arrangement to its
restoring response, rather than inferring dynamical stability from
energy alone. An imaginary harmonic mode identifies a collective
displacement along which the calculated energy decreases; imaginary
frequencies are plotted as negative values in this work.

The minimum sampled frequency is retained as a continuous quantity.
Structures are classified as stable for \(\omega_{\min}\geq-0.10\) THz,
borderline for \(-0.30\leq\omega_{\min}<-0.10\) THz, and unstable for
\(\omega_{\min}<-0.30\) THz. These margins separate nearly soft modes
from pronounced harmonic instabilities. Only structures with converged
relaxations and finite phonon results enter the 2,649,640-structure
dataset.

Each relaxed parent's EquFlashV2 energy is placed on a fixed reference
hull compatible with the Materials Project 2020 (MP2020)
corrections\textsuperscript{\citeproc{ref-wang2021corrections}{46}}.
This model-derived hull estimate is distinct from the source values and
describes the same atomic arrangement as the phonon frequencies.

\FloatBarrier

\subsection{3. Data Record}\label{data-record}

This construction yields 2,649,640 structures, comprising 2,516,445
retained source-derived structures and 133,195 explicitly generated
inversion partners, with 31,997,598 atoms in their primitive cells.
Among the full collection, 265,664 structures are identified as chiral
in their source cells. The paired Te structures in Fig. 1c illustrate
why opposite handed geometries are retained: spatial inversion preserves
crystal energies and phonon frequencies, but can reverse
handedness-dependent properties. Generated partners retain inverted
coordinates and inherit their parents' scalar energies and phonon
frequencies. Supplementary Figure S1a-c resolves these populations
across enantiomorphic pairs and other Sohncke space groups.

The same structural summary spans 89 elements, 219 space groups, and
primitive cells containing 1-48 atoms. Its eight relaxed examples
progress from Si and NaCl through PrAgHg and ZrCdInHg to
oxygen-containing compounds with five to eight elements. All eight
satisfy the adopted stable-class criterion.

Parquet tables store each atomic arrangement as lattice vectors,
species, and fractional coordinates in JavaScript Object Notation
(JSON). Each entry links source and relaxed-structure hull energies to
the minimum phonon frequency and stability class. Units, calculation
conditions, source attribution, and inversion partners accompany these
quantities. Supplementary Methods S3 defines the quantities, units, and
populations underlying the reported proportions.

Data repository: Zenodo (version 1.0.0). Permanent identifier:
\href{https://doi.org/10.5281/zenodo.23572554}{\textcolor{black}{10.5281/zenodo.23572554}}. Reviewer access: \href{https://zenodo.org/records/23572554}{\textcolor{black}{public Zenodo record 23572554}}.

\subsection{4. Data Overview}\label{data-overview}

\subsubsection{4.1 Diversity of the
dataset}\label{diversity-of-the-dataset}

Crystoria brings lattice-dynamical information to a chemically broad
collection spanning 89 elements. Substantial populations extend across
the main-group, transition-metal, and lanthanide regions of the periodic
table (Fig. 2a). Oxygen is the most frequent element, occurring in
271,353 structures, while La and Tb each occur in more than 190,000. The
chemical coverage therefore includes both light-element compounds and
large populations containing heavier metals, placing different atomic
environments within the same dynamical-stability study.

At the compound level, this breadth is concentrated in multicomponent
crystals. The composition-order distribution peaks at three elements and
remains high at four; together, ternary and quaternary compounds
contribute 2,400,530 structures (90.60\%; Fig. 2b). The population then
falls steeply, but 82,193 structures contain five to eight elements. The
principal chemical scope of Crystoria is therefore not the elemental or
binary limit, but multicomponent lattices in which several atomic
species contribute to the collective vibrational response. The
higher-order tail extends this coverage to more compositionally complex
compounds.

\begin{figure}[!htbp]
\centering
\includegraphics[width=\linewidth,height=0.67\textheight,keepaspectratio]{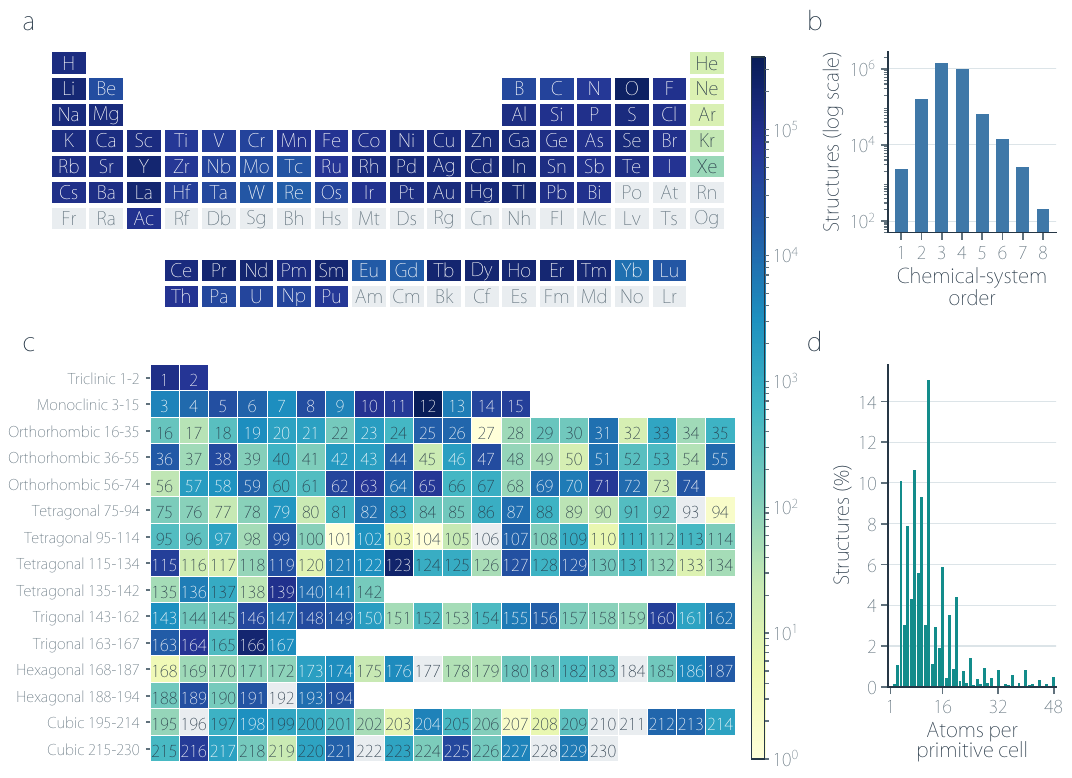}
\caption{Chemical and structural diversity of Crystoria. All panels describe the
same 2,649,640 structures. (a) Number of structures containing each
element; a compound contributes to every element it contains. (b) Number
of distinct elements per structure, from one to eight, with structure
counts on a logarithmic scale. (c) Distribution over the 230 space
groups, with 219 represented. These symmetry assignments refer to the
source-standardized cells. (d) Number of atoms in the primitive cell
over the 1--48-atom domain. Panels (a) and (c) share a logarithmic count
scale; gray space-group cells indicate absence. Space-group numbers
appear within the cells.}
\label{fig:2}
\end{figure}

The variety of compositions is accompanied by broad crystallographic
coverage: 219 of the 230 space groups and all seven crystal systems are
represented (Fig. 2c). The widely varying tile intensities distinguish
dense symmetry families from sparsely populated groups. Space group 12
contains the largest population, with 378,575 structures (14.29\%), and
substantial populations also occupy tetragonal and trigonal groups. The
database thus extends well beyond a collection of familiar cubic
lattices. Because crystal symmetry constrains the equivalence of atomic
sites and displacements, this coverage makes lattice-dynamical
information available across distinct forms of crystallographic order,
not only across different compositions.

This combined chemical and crystallographic diversity is mostly
accommodated within compact primitive cells. The cell-size distribution
has pronounced peaks at small atom counts, with the largest at 12 atoms
(399,001 structures), followed by a less populated tail extending to 48
atoms (Fig. 2d). Overall, 2,418,571 structures (91.28\%) contain at most
20 atoms per primitive cell; this fraction reaches 93.00\% among ternary
and quaternary compounds. Multicomponent chemistry is therefore not
confined to large cells. Taken together, the four distributions identify
Crystoria as a chemically and crystallographically diverse collection
whose main population comprises compact multicomponent crystals, with
additional coverage of larger atomic arrangements.

\FloatBarrier

\subsubsection{4.2 Stability of the
dataset}\label{stability-of-the-dataset}

The source-to-relaxed hull-energy map contains a dense ridge near the
equality line, indicating broad agreement between the two energy
descriptions, together with a spread on both sides (Fig. 3a). In total,
23.51\% of structures differ by more than 20 meV atom\(^{-1}\). These
shifts combine changes in atomic geometry with differences in the energy
model, corrections, and reference phases. Supplementary Figure S2
resolves median absolute differences by element, space group, and cell
size for independent source-derived structures. Re-evaluating the energy
of the relaxed geometry therefore matters even when the starting
structures have already passed a low-energy selection.

The relaxed-structure distribution nevertheless remains strongly
concentrated near the reference hull. Its cumulative distribution rises
rapidly at low energies, with a median of 21.84 meV atom\(^{-1}\) and a
95th percentile of 87.93 meV atom\(^{-1}\). In the full collection,
2,591,836 structures (97.82\%) remain within 100 meV atom\(^{-1}\) of
the reference hull. Thus, within the common model description, the
expanded structural coverage retains a predominantly low-energy
character.

\begin{figure}[!htbp]
\centering
\includegraphics[width=\linewidth,height=0.67\textheight,keepaspectratio]{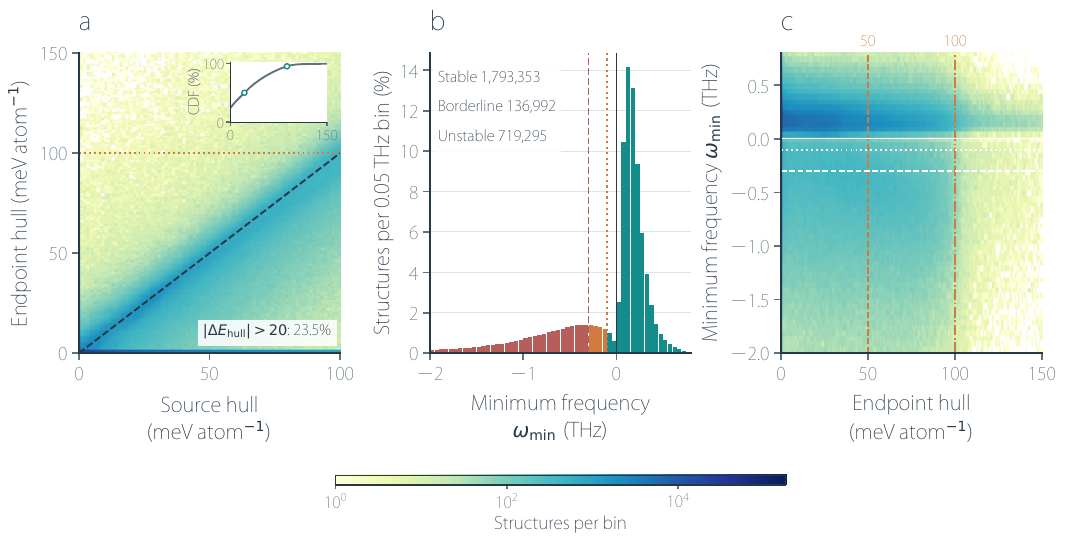}
\caption{Thermodynamic and dynamical stability landscape of Crystoria. All panels
use 2,649,640 structures with converged relaxations and finite phonon
results. (a) Source and relaxed-structure hull energies; 18,209 values
above 150 meV atom\(^{-1}\) are outside the map. The diagonal marks
equality; the fraction differing by more than 20 meV atom\(^{-1}\) is
given within the panel. The inset gives the cumulative distribution of
relaxed-structure hull energies, with median and 95th-percentile values
of 21.84 and 87.93 meV atom\(^{-1}\). (b) Minimum phonon frequency in
0.05 THz bins, colored by stability class; 129,067 values lie outside
the display. Imaginary modes are represented by negative frequencies.
(c) Joint energy--frequency distribution containing 2,503,993 structures
in the displayed range. Vertical references mark 50 and 100 meV
atom\(^{-1}\); horizontal references mark \(-0.30\), \(-0.10\), and 0
THz. Panels (a) and (c) share a logarithmic count scale. The 133,195
generated inversion partners have inherited scalar values. Hull energies
are MP2020-aligned EquFlashV2 estimates for the relaxed structures.}
\label{fig:3}
\end{figure}

The minimum-frequency distribution separates this low-energy population
into a sharp peak on the positive-frequency side of zero and a broad
tail extending into imaginary frequencies (Fig. 3b). Under the stated
margins, 1,793,353 structures (67.68\%) are stable, 136,992 (5.17\%) are
borderline, and 719,295 (27.15\%) are unstable. Crystoria therefore
provides a million-scale population satisfying a common
harmonic-stability criterion, together with nearly soft and unstable
lattices that reveal where that criterion is not met. The continuous
frequencies distinguish proximity to the stability margin from
pronounced imaginary modes.

The unstable population also clarifies why structural relaxation cannot
replace a phonon calculation. Although 2,624,127 structures (99.04\%)
preserve their source space-group type under relaxation, 27.07\% of this
symmetry-preserving population is dynamically unstable. The full
group-to-group and crystal-system comparisons use common symmetry
tolerances (Supplementary Figure S4). The broad imaginary-frequency tail
therefore does not arise only from crystals that change symmetry. Small
residual forces constrain the first derivatives of energy, whereas
phonons probe its curvature under collective displacements. A relaxed
lattice can retain its symmetry while remaining susceptible to an
energy-lowering distortion.

The overall stable fraction also conceals substantial differences
between chemical and crystallographic populations. Within Crystoria,
38.51\% of oxygen-containing structures satisfy the stable-class
criterion, compared with 71.01\% of oxygen-free structures. Stable-class
fractions are 31.92\% in space group 1 and 73.05\% in space group 12.
The tendency to develop imaginary modes is therefore uneven across the
crystal landscape, while stable and unstable structures coexist even
within a fixed space-group type. A common phonon calculation resolves
differences that composition or symmetry alone cannot determine for an
individual crystal.

Thermodynamic and dynamical stability are associated but not
interchangeable. A dense positive-frequency band extends over a broad
range of hull energies, while imaginary modes remain populated close to
the hull (Fig. 3c). Stable-class structures have a lower median hull
energy than unstable structures, 16.76 versus 37.65 meV atom\(^{-1}\),
consistent with an overall association between favourable energetics and
lattice stability. Yet 187,983 of the 980,560 structures within 10 meV
atom\(^{-1}\) of the reference hull remain dynamically unstable
(19.17\%). Even a stringent energetic selection therefore retains many
crystals whose atomic arrangement is unstable against a collective
distortion.

The converse population is also substantial: 339,576 stable-class
structures lie more than 50 meV atom\(^{-1}\) above the reference hull.
These lattices satisfy the adopted harmonic-stability criterion while
remaining energetically unfavourable relative to competing phase
combinations. They are metastable candidates with a different physical
character from near-hull crystals with imaginary modes. These
contrasting populations expose the distinction between a driving force
toward competing phases and an unstable displacement of a particular
lattice.

Requiring both a relaxed-structure hull energy no greater than 50 meV
atom\(^{-1}\) and the stable-class phonon margin selects 1,453,777
structures (54.87\% of Crystoria). This joint population identifies
candidates that combine energetic proximity to competing phases with the
adopted harmonic-stability criterion. The value of the database is
therefore not only its large stable subset, but its ability to resolve
energetic competition and lattice response separately, and then select
their intersection across a broad crystal-structure space.

\FloatBarrier

\subsection{5. Technical Validation}\label{technical-validation}

We assess the model-derived phonon classifications against 9,958
nonmagnetic semiconductors with DFT phonons calculated using the
Perdew--Burke--Ernzerhof (PBE)
approximation\textsuperscript{\citeproc{ref-Loew2025Phonons}{35}}. Both
calculations use the same fixed DFT structures, supercells, and
reciprocal-space points, isolating the predicted vibrational response
from differences in model relaxation (Supplementary Methods S2).
EquFlashV2 gives 95.11\% accuracy and 91.22\% balanced accuracy, the
mean recall of the two reference classes. The 95\% confidence interval
for balanced accuracy is 90.36--92.06\%, estimated by resampling groups
of identical crystal geometries. Supplementary Figure S3 gives the
reference populations, class-resolved errors, model comparisons, and
uncertainty estimates for both the full and balanced reference sets.

Spatial inversion leaves crystal energies and phonon frequencies
unchanged, providing a symmetry-based consistency test. The phonon
classes agree in 208 of 209 independently calculated inversion pairs,
with a maximum minimum-frequency difference of 0.05444 THz.
Supplementary Figure S1d compares the independently calculated energies
and phonon frequencies of these pairs.

\subsection{6. Usage Notes}\label{usage-notes}

Crystoria can support the training and fine-tuning of crystal-generation
models using structures that satisfy the stated harmonic-stability
criterion. Conditional generation could use the relaxed-structure hull
estimate and minimum phonon frequency as distinct targets, linking phase
competition and lattice response in the design objective.

Stable, borderline, and unstable examples can also support predictors
that guide generation or rank candidates. Such predictors could use
continuous frequencies to distinguish nearly soft lattices from
pronounced harmonic instabilities.

Temperature-dependent studies involve further calculations of
vibrational free energies and, where needed, anharmonic
effects\textsuperscript{\citeproc{ref-Togo2015Phonons}{22}}. Magnetic
configurations have not been systematically enumerated and must be
specified for magnetic-state comparisons.

\subsection{Data Availability}\label{data-availability}

Crystoria comprises the 2,649,640 structures and associated energetic
and dynamical quantities described in Data Record. Repository:
Zenodo (version 1.0.0). Persistent identifier: \href{https://doi.org/10.5281/zenodo.23572554}{\textcolor{black}{10.5281/zenodo.23572554}}. Newly generated Crystoria data are released under
CC BY 4.0; third-party data retain their source-specific terms and
attribution.

\subsection{Code Availability}\label{code-availability}

Structure-curation, phonon-calculation and analysis programs: \href{https://github.com/KeXin-Pang/Crystoria/releases/tag/v1.0.0}{\textcolor{black}{GitHub release v1.0.0}}. Persistent archival identifier: \href{https://doi.org/10.5281/zenodo.23572554}{\textcolor{black}{10.5281/zenodo.23572554}}. Original Crystoria method code is released under
the MIT License. Third-party code and the separately identified GGNN
compatibility patches retain their original licenses, including CC
BY-NC-SA 4.0 for GGNN.

\subsection{Supplementary Information}\label{supplementary-information}

The accompanying Supplementary
Information\textsuperscript{\citeproc{ref-CrystoriaSupplementaryInformation}{44}}
contains the construction methods, reliability tests, physical-quantity
definitions, and Supplementary Figures S1-S4.

\FloatBarrier

\subsection{References}\label{references}

\protect\phantomsection\label{refs}

\subsection{Author Contributions}\label{author-contributions}

Ke-Xin Pang performed the main calculations and drafted the manuscript.
Ke-Xin Pang and Yan Gao conceived the study. Ze-Feng Gao, Yan Gao, and
Zhong-Yi Lu validated the results, supervised the research, and acquired
funding. All authors discussed the results and revised the manuscript.

\subsection{Competing Interests}\label{competing-interests}

The authors declare no competing interests.

\subsection{Funding}\label{funding}

This work was supported by the National Natural Science Foundation of
China (Grant Nos. 12304202, 62476278, and 12434009), the Beijing Natural
Science Foundation (Grant Nos. Z260002 and Z250005), the National Key
R\&D Program of China (Grant No.~2024YFA1408601), the Hebei Natural
Science Foundation (Grant No.~A2023203007), and the Science Research
Project of Hebei Education Department (Grant No.~BJK2024085).

\subsection{Disclosure of AI Use}\label{disclosure-of-ai-use}

The authors used AI for language polishing of the manuscript.

\end{document}


\begin{center}
{\Large\bfseries Supplementary Information\par}
\end{center}

Crystoria: A Large-Scale Crystal Dataset Linking Thermodynamic and
Dynamical Stability for Materials AI

Ke-Xin Pang\(^{1}\), Ze-Feng Gao\(^{2,3,*}\), Yan Gao\(^{1,\dagger}\),
and Zhong-Yi Lu\(^{2,3,\ddagger}\)

\subsection{Supplementary Methods S1. Crystal Structures and Lattice
Dynamics}\label{supplementary-methods-s1.-crystal-structures-and-lattice-dynamics}

\subsubsection{S1.1 Crystal collection and physical
domain}\label{s1.1-crystal-collection-and-physical-domain}

Crystoria contains 2,649,640 ordered, three-dimensional crystal
structures, with 31,997,598 atoms in their primitive cells. The selected
cells contain 1--48 atoms, have supported atomic species with
\(Z\leq100\), and have minimum interatomic separations of at least 0.50
Å. Source attribution and the conditions of the original thermodynamic
calculations accompany each structure.

The contributing collections are Alexandria, Materials Project, JARVIS,
NIMS HeuslerDB, and OQMD. Some Alexandria and Materials Project data are
obtained through the collections distributed with MatterGen, including
its DFT relaxation and energy
processing\textsuperscript{\citeproc{ref-Zeni2025MatterGen}{1}}. A
crystal appearing in several collections remains associated with its
original sources.

Energy above the convex hull measures competition with other phases at
the same overall composition. The source-energy selection requires every
retained admissible estimate to be no greater than 100 meV
atom\(^{-1}\). The minimum and maximum admissible source estimates are
retained separately, and the maximum controls the initial selection.
This cutoff is applied to source estimates before relaxation. The
source-to-relaxed comparison in Fig. 3 uses the source hull energy
associated with each structure's original calculation.

\subsubsection{S1.2 Equivalent cells and independent
comparisons}\label{s1.2-equivalent-cells-and-independent-comparisons}

Different lattice bases, origins, or atomic orderings can describe the
same periodic crystal. Primitive-cell reduction and symmetry analysis
use pymatgen and
spglib\textsuperscript{\citeproc{ref-Ong2013Pymatgen}{2},\citeproc{ref-togo2024spglib}{3}}.
Initial symmetry processing uses a 0.01 Å positional tolerance and a
\(5^\circ\) angular tolerance, with fractional coordinates rounded to
six decimal places.

Only structures with the same composition are compared. Candidate pairs
are selected using reduced lattice lengths, angles, volume, and periodic
interatomic distances. The allowed differences are 0.001 for sorted
logarithmic lattice lengths and sorted absolute angle cosines, and 0.003
for logarithmic volume. Distance percentiles at 0, 10, 25, 50, 75, 90,
and 100\% are normalized by \((V/N)^{1/3}\), where \(V\) is cell volume
and \(N\) is atom count; their comparison window is 0.0025.

Equivalence requires a direct, species-preserving correspondence between
periodic atomic positions. Relative lattice and normalized
atomic-position tolerances are \(10^{-4}\), with a \(0.01^\circ\) angle
tolerance. The comparison permits proper rotations but not reflection,
lattice scaling, additional primitive reduction, or supercell matching.
Representative selection does not depend on the phonon class or other
property values.

\subsubsection{S1.3 Relaxed structures and thermodynamic
quantities}\label{s1.3-relaxed-structures-and-thermodynamic-quantities}

Both the cell and atomic positions are relaxed with the same EquFlashV2
interatomic potential, using a maximum-force criterion of 0.03 eV
Å\(^{-1}\) and at most 200 steps. The model is the 45-million-parameter
version compared with density-functional theory (DFT) in Figure S3.

For the relaxed structures, EquFlashV2 single-point energies are
corrected consistently with the Materials Project 2020 (MP2020) scheme
and compared with a fixed reference phase
diagram\textsuperscript{\citeproc{ref-wang2021corrections}{4}}. Both
signed and nonnegative distances from this reference hull are retained
as outputs of the common model calculation. Source and relaxed-structure
values remain attached to the atomic arrangements from which they were
obtained. Their differences combine relaxation with changes in the
interatomic potential, energy corrections, and reference phases. Figure
S2 shows how these differences vary with composition, symmetry, and cell
size.

\subsubsection{S1.4 Phonon calculations and stability
margins}\label{s1.4-phonon-calculations-and-stability-margins}

The harmonic response is calculated with Phonopy using atomic
displacements of 0.01 Å in \(2\times2\times2\)
supercells\textsuperscript{\citeproc{ref-Togo2015Phonons}{5}}. The mean
force is removed from each displaced supercell, and the reconstructed
force constants are symmetrized. Force changes are evaluated without
separately subtracting the forces of the undisplaced supercell.

Frequencies are evaluated on a non-\(\Gamma\)-centred
\(12\times12\times12\) reciprocal-space mesh. We retain the uncorrected
minimum over this mesh as a continuous quantity, with imaginary modes
represented by negative values. A negative harmonic mode identifies an
atomic displacement along which the calculated energy decreases near the
relaxed structure.

The reported classes are stable for \(\omega_{\min}\geq-0.10\) THz,
borderline for \(-0.30\leq\omega_{\min}<-0.10\) THz, and unstable for
\(\omega_{\min}<-0.30\) THz. These margins separate pronounced harmonic
instabilities from nearly soft modes.

All structures included in the dataset have finite phonon results and
converged parent relaxations.

\subsubsection{S1.5 Inversion partners and
chirality}\label{s1.5-inversion-partners-and-chirality}

Spatial inversion generates the missing handed partner through
\(\mathbf{x}\mapsto-\mathbf{x}\pmod 1\), with the lattice and species
order retained. Existing partners are connected rather than added again.
The chirality assessment uses consistent source-cell symmetry
assignments within the 65 Sohncke space-group types.

The dataset contains 133,195 explicitly generated inversion partners.
They inherit the parent energy and phonon results as parity-even scalar
quantities, which are unchanged by spatial inversion.
Handedness-dependent quantities, such as optical-rotation sign or
topological charge, are not inherited. Chirality counts refer to the
source-cell symmetry, while phonon properties refer to the relaxed
arrangements. Independently calculated inversion pairs provide a
separate consistency test in Supplementary Methods S2.2 and Figure S1.

\subsection{Supplementary Methods S2. Reliability and Physical
Interpretation}\label{supplementary-methods-s2.-reliability-and-physical-interpretation}

\subsubsection{S2.1 Crystal symmetry before and after
relaxation}\label{s2.1-crystal-symmetry-before-and-after-relaxation}

Space-group types are determined independently from the source and
relaxed atomic coordinates using spglib, with a common positional
tolerance of 0.01 Å and an angular tolerance of \(5^\circ\). All
2,649,640 structures enter this comparison, including generated
inversion partners.

The space-group type is unchanged for 2,624,127 structures (99.04\%) and
changes for 25,513 (0.96\%). Agreement ranges from 95.82\% for triclinic
source structures to 99.95\% for cubic source structures. Figure S4
gives the complete correspondence over space groups 1--230 and the
population and agreement within each crystal system. Among structures
retaining their space-group type, 710,222 (27.07\%) are dynamically
unstable. Symmetry retention and harmonic stability therefore describe
different aspects of the relaxed lattice.

\subsubsection{S2.2 Inversion symmetry}\label{s2.2-inversion-symmetry}

Crystal energies and harmonic phonon frequencies are unchanged by
spatial inversion. Among 209 independently calculated inversion pairs,
208 have the same phonon class. The largest minimum-frequency difference
is 0.05444 THz. Figure S1 shows the corresponding symmetry populations
and independent pair comparisons.

\subsubsection{S2.3 Comparison with DFT
phonons}\label{s2.3-comparison-with-dft-phonons}

The external reference comprises 9,958 nonmagnetic semiconductors with
DFT phonon calculations using the Perdew--Burke--Ernzerhof (PBE)
approximation: 8,189 stable and 1,769
unstable\textsuperscript{\citeproc{ref-Loew2025Phonons}{6}}. EquFlashV2
and MatterSim forces are evaluated on identical 0.01 Å displacements of
the fixed DFT structures. The original supercell matrices and
commensurate reciprocal-space points are retained. Mean-force removal
and force-constant reconstruction are common to both models, and no
model relaxation is applied.

The reference uses a numerical tolerance of \(-35\) K for acoustic modes
at \(\Gamma\). This reference convention differs from the frequency
margins and fixed reciprocal-space mesh used for Crystoria.

EquFlashV2 gives 95.11\% accuracy and 91.22\% balanced accuracy on the
full reference set. Balanced accuracy is the mean recall of the stable
and unstable classes, avoiding dominance by the more numerous stable
structures. Stable recall is 97.25\%, unstable recall is 85.19\%, and
stable precision is 96.81\%.

The 95\% confidence interval for balanced accuracy is 90.36--92.06\%,
obtained by resampling groups of identical crystal geometries.
Differences between the two potentials are assessed with a paired exact
McNemar test. Figure S3 also includes a balanced set of 2,688 structures
with agreeing PBE and PBEsol classifications, and reports the model
differences and their uncertainty. Overlap with the training data of
either potential remains undetermined.

\subsection{Supplementary Methods S3. Physical Quantities and Use of the
Data}\label{supplementary-methods-s3.-physical-quantities-and-use-of-the-data}

\subsubsection{S3.1 Structural and energetic
information}\label{s3.1-structural-and-energetic-information}

Each structure is described by its lattice vectors, atomic species, and
fractional coordinates, together with composition, primitive-cell atom
count, and source symmetry. The machine-readable tables use Parquet,
with atomic structures represented in JavaScript Object Notation (JSON).

Thermodynamic information distinguishes source hull energies from the
estimate for the relaxed structure. Dynamical quantities comprise the
sampled minimum frequency and its stability class. Energies are reported
per atom and frequencies in THz, with source and calculation conditions
accompanying the values.

The chemical, symmetry-resolved, and energy-conditioned proportions in
the main text use the complete 2,649,640-structure population. Chemical
and symmetry-resolved populations use the source compositions and space
groups shown in Figure 2; energy windows use hull estimates for the
relaxed structures. Generated inversion partners remain included in
these population statistics.

\clearpage
\begin{figure}[!htbp]
\centering
\includegraphics[width=\linewidth,height=0.67\textheight,keepaspectratio]{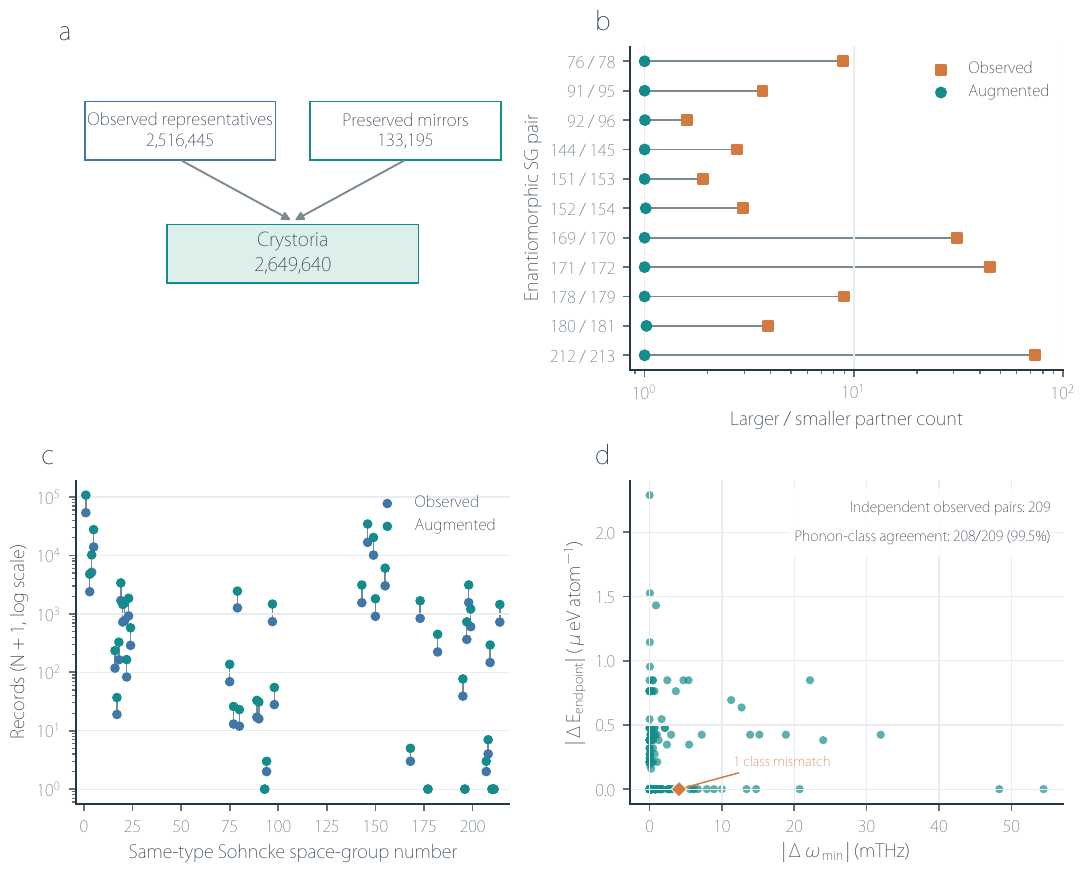}
\caption{Distinct crystal structures and inversion partners. All panels describe
the same 2,649,640 structures. (a) The population comprises 2,516,445
source-derived structures and 133,195 generated inversion partners. (b)
Relative populations of the two partners in each of the 11
enantiomorphic space-group pairs, before and after adding missing
inversion partners. Ratios use observed counts; a ratio is omitted if
its denominator is zero. (c) Populations of the other 43 Sohncke
space-group types, with 38 represented. Counts are plotted as \(N+1\) on
a logarithmic scale. Symmetry assignments in (b,c) refer to the source
structures. (d) Differences in relaxed energy and minimum phonon
frequency for 209 independently calculated inversion pairs. The class
agrees in 208 pairs (99.5\%). Generated partners inherit scalar values
from their calculation parents. Structural comparison criteria are given
in Supplementary Methods S1.2.}
\label{fig:S1}
\end{figure}

\clearpage
\begin{figure}[!htbp]
\centering
\includegraphics[width=\linewidth,height=0.67\textheight,keepaspectratio]{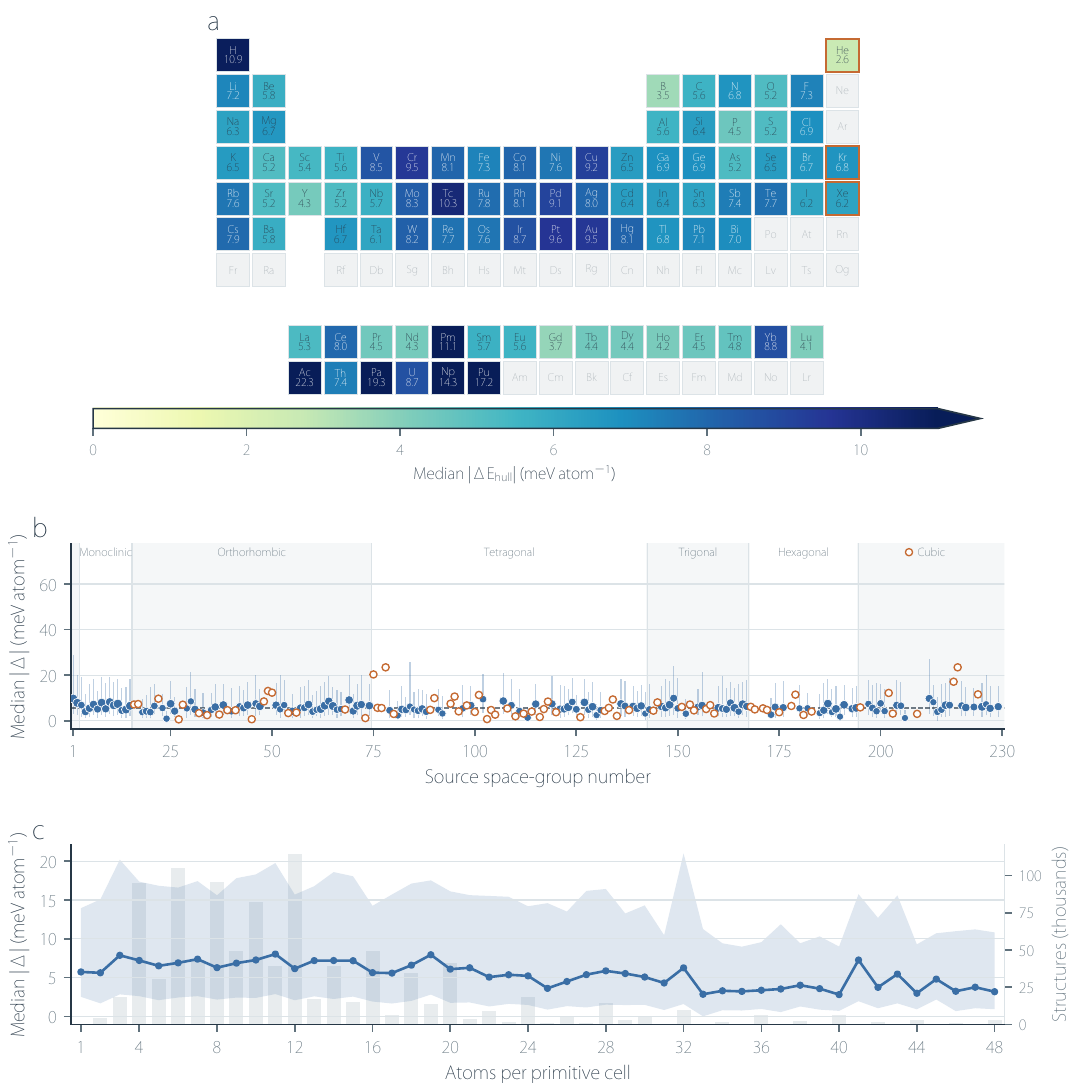}
\caption{Comparison of source and relaxed-structure hull energies. Absolute
differences compare the selected source energy above the convex hull
with the nonnegative MP2020-aligned EquFlashV2 estimate for the relaxed
structure. The comparison includes 993,629 independent source-derived
structures, excluding generated inversion partners and cases using a
substitute reference-hull estimate. (a) Median absolute difference for
each of 87 represented elements. A compound contributes to each
constituent element; orange borders indicate fewer than 100 structures.
(b) Median difference for the 217 represented source space groups, with
interquartile ranges. Filled points represent at least 100 structures
and open points fewer than 100. (c) Median difference as a function of
primitive-cell atom count, with interquartile ranges shaded and sample
counts in gray. Differences reflect changes in geometry, potential,
energy corrections, and reference phases.}
\label{fig:S2}
\end{figure}

\clearpage
\begin{figure}[!htbp]
\centering
\includegraphics[width=\linewidth,height=0.67\textheight,keepaspectratio]{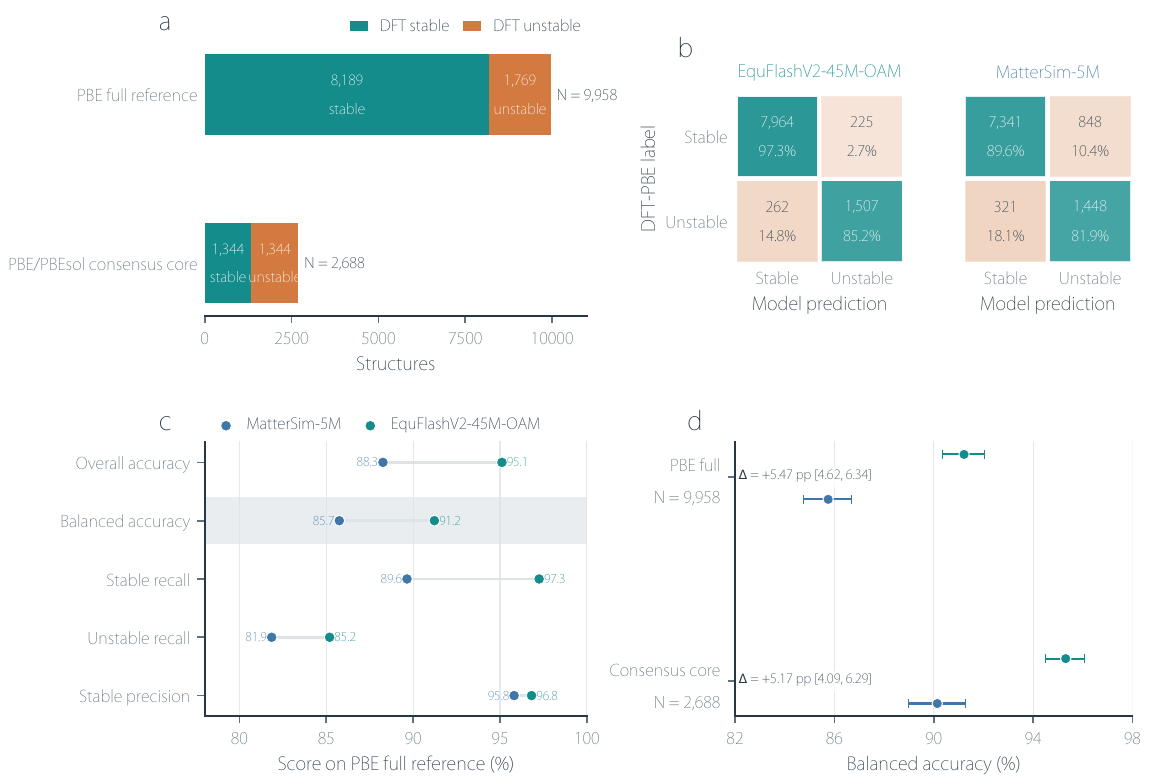}
\caption{Phonon stability classifications compared with DFT. (a) Stable and
unstable populations in the complete PBE reference set of 9,958
structures (8,189 stable; 1,769 unstable) and the balanced
PBE/PBEsol-consensus set of 2,688 structures. (b) Confusion matrices
normalized within each reference class, with structure counts shown. (c)
Classification measures on the complete reference set. EquFlashV2 gives
95.11\% accuracy and 91.22\% balanced accuracy, compared with 88.26\%
and 85.75\% for MatterSim. (d) Balanced accuracy and 95\% confidence
intervals obtained by resampling groups of identical crystal geometries.
EquFlashV2 exceeds MatterSim by 5.47 percentage points on the full set
(95\% interval 4.62--6.34; paired exact McNemar
\(p=2.93\times10^{-116}\)) and 5.17 points on the balanced set
(4.09--6.29; \(p=9.43\times10^{-21}\)). Both potentials use identical
0.01 Å displacements of fixed DFT structures, the source supercell
matrices and commensurate wavevectors, mean-force removal, and common
force-constant reconstruction. Reference classes and predictions use the
source \(-35\) K acoustic-mode tolerance at \(\Gamma\).}
\label{fig:S3}
\end{figure}

\clearpage
\begin{figure}[!htbp]
\centering
\includegraphics[width=\linewidth,height=0.67\textheight,keepaspectratio]{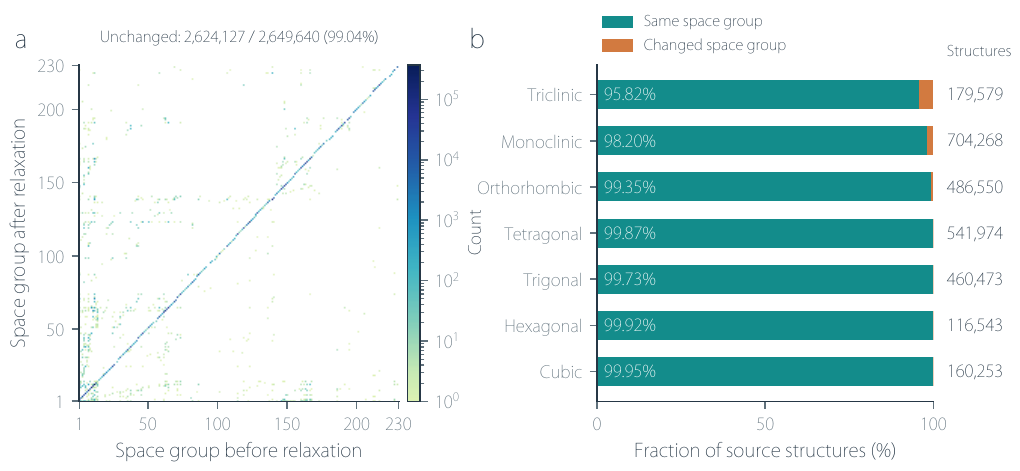}
\caption{Space-group correspondence before and after relaxation. Both symmetries
are determined independently for all 2,649,640 structures, including
generated inversion partners, using a common 0.01 Å positional tolerance
and \(5^\circ\) angular tolerance. (a) Full correspondence between
source and relaxed space-group types 1--230. Color gives the structure
count on a logarithmic scale; white cells contain no structures. The
diagonal contains the 2,624,127 structures (99.04\%) retaining their
space-group type; the remaining 25,513 (0.96\%) lie off the diagonal.
(b) Fractions retaining or changing space-group type within each source
crystal system. Percentages inside the bars indicate agreement, and
numbers at right give each system's full population. Matching
space-group types do not require identical lattice parameters or atomic
positions.}
\label{fig:S4}
\end{figure}

\subsection{Supplementary References}\label{supplementary-references}

\protect\phantomsection\label{refs}